\documentclass{ws-rv961x669}
\usepackage[square]{ws-rv-van}     
\usepackage{ws-rv-thm}     
\usepackage{subfigure}     
\makeindex

\begin{document}

\chapter[Sum Rules for Leptons]{Sum Rules for Leptons}\label{ra_ch1}

\author[M.~Spinrath]{Martin~Spinrath}

\address{
Institut f\"ur Theoretische Teilchenphysik, Karlsruhe Institute of Technology,\\
Engesserstra\ss{}e 7, D-76131 Karlsruhe, Germany\\
E-mail: martin.spinrath@kit.edu\\
www.ttp.kit.edu}

\begin{abstract}
There is a wide class of models which give a dynamical description of the
origin of flavour in terms of spontaneous symmetry breaking of an underlying
symmetry. Many of these models exhibit sum rules  which relate
on the one hand mixing angles and the Dirac CP phase with each other
and/or on the other hand neutrino masses and Majorana phases with each other.
We will briefly sketch how this happens and discuss briefly the impact of
renormalisation group corrections to the mass sum rules.
\end{abstract}


\body


\section{Introduction}

The origin of flavour in the Standard Model of Particle Physics (SM) is still
a big puzzle. It is not clear why there are three generations of fermions
exhibiting this very peculiar patterns of masses and mixing parameters.
A very popular approach in recent years has been the use of non-Abelian
(discrete) family symmetries driven by the rather large mixing angles in the lepton
sector, for recent reviews, see, e.g., \cite{Ishimori:2010au}.

Nevertheless, in this proceedings we do not want to dive into cumbersome
model building details. Instead we want to focus on two classes of
predictions which appear in a very wide class of flavour models. To be precise
we want to discuss two kinds of sum rules. The first type, mixing sum rules,
relates the leptonic mixing angles to the Dirac CP violating phase while
while the second type, mass rum rules, relates the neutrino masses to
the Majorana phases.

After discussing the two cases separately we will give an example where both
kinds of sum rules appear which makes the model extremely predictive. And then
we will summarise and conclude.

\section{A Short Note on Mixing Sum Rules}

Probably better known and studied then the mass sum rules are so-called mixing sum rules.
Typically, they emerge when the effective light Majorana neutrino mass matrix exhibits a
symmetry pattern, like, for instance, bimaximal mixing \cite{Petcov:1982ya} or tri-bimaximal
mixing \cite{Xing:2002sw}. If the charged lepton mass matrix would be diagonal in that basis,
the leptonic mixing angles would be exactly predicted at the symmetry breaking scale. But since
many of the popular mixing schemes exhibited a vanishing reactor mixing angle $\theta_{13} =0$,
this setup is disfavoured by the measurement of $\theta_{13} \approx 9^\circ$.

There are plenty of possible modifications on the market which we cannot all
discuss here exhaustively. Instead, we chose a case which we consider to be
well motivated. Namely, that the charged lepton mass matrix is not simply
diagonal, but has a sizeable mixing of the order of the Cabibbo angle for the first
two generations. This is exactly what one would expect in a grand unified setup
where the Yukawa matrices of the leptons are related to the quark Yukawa couplings.

Then one would find, for instance, for bimaximal mixing
\begin{equation}
\sin^2 \theta_{12} \approx \frac{1}{2} + \sin \theta_{13} \cos \delta
\end{equation}
and for tri-bimaximal mixing
\begin{equation}
\sin^2 \theta_{12} \approx \frac{1}{3} +  \frac{2 \sqrt{2}}{3} \sin \theta_{13} \cos \delta
\end{equation}
to leading order in the $\theta_{13}$ expansion, see, for instance, \cite{Marzocca:2011dh}
and references therein.

Nowadays all mixing angles have been measured such that these sum rules
can be translated into constraints on the Dirac CP violating phase $\delta$.
For the bimaximal case CP should be almost conserved($\cos \delta \approx -1$)
while for the tri-bimaximal case CP should be strongly violated ($\cos \delta \lesssim 0$).

\section{Neutrino Mass Sum Rules}

Neutrino mass sum rules emerge somewhat accidental in flavour models. They are not
related to a special family symmetry or a subgroup thereof. They are also not specific to
any seesaw mechanism. The reason for them is simply that due to a very economic breaking
of the family symmetry it can happen that the three light complex Majorana neutrino masses
depend effectively on two complex parameters only. From simply counting the degrees of freedom
it is clear that there should be two relations which can be expressed as a complex sum rule
for the neutrino masses including the Majorana phases.

For instance in the $SU(5) \times A_5$ flavour model in \cite{Gehrlein:2014wda} we had a type I seesaw
mechanism where the neutrino Yukawa matrix $Y$ and the right neutrino mass matrix $M_{RR}$
had the structures
\begin{equation}
Y \sim \begin{pmatrix} 1 & 0 & 0 \\ 0 & 0 & 1 \\ 0 & 1 & 0 \end{pmatrix} \text{ and }
M_{RR} \sim \begin{pmatrix} 
 2 \sqrt{\frac{2}{3}} (v_2 + v_3) & - \sqrt{3} v_2 & - \sqrt{3} v_2 \\
 - \sqrt{3} v_2 & \sqrt{6} v_3 & - \sqrt{\frac{2}{3}} (v_2 + v_3) \\
 - \sqrt{3} v_2 & - \sqrt{\frac{2}{3}} (v_2 + v_3) & \sqrt{6} v_3
 \end{pmatrix} \;,
\end{equation}
where $v_2$ and $v_3$ are complex flavon vacuum expectation values
which break the family symmetry $A_5$.
Then it is obvious that the light neutrino mass matrix will depend only on two
effective parameters and in fact in this case we find the mass sum rule
\begin{equation}
\frac{\text{e}^{\text{i } \phi_1}}{m_1} + \frac{\text{e}^{\text{i } \phi_2}}{m_2} = \frac{1}{m_3}  \;,
\end{equation}
where $m_i$, $i = 1,2,3$, are the three light neutrino masses and $\phi_1$
and $\phi_2$ the Majorana phases.

But this is not the only known sum rule in the literature. We found in total twelve
different sum rules which can all be parametrised in the following form
\begin{equation}
 s \equiv c_1 \left( m_1 \text{e}^{\text{i } \phi_1} \right)^d \text{e}^{\text{i } \Delta \chi_{13}}
  + c_2 \left( m_2 \text{e}^{\text{i } \phi_2} \right)^d \text{e}^{\text{i } \Delta \chi_{23}}
  + m_3^d \stackrel{!}{=} 0 \;,
\end{equation}
where $c_i$, $\Delta \chi_{ij}$ and $d$ are given by the underlying model but can
only take discrete values. In the previous example, for instance, $c_1 = c_2 =1$,
$d = -1$ and $\Delta \chi_{13} = \Delta \chi_{23} = \pi$. A complete list of the sum rules
we found in the literature is given in Table~\ref{tab:overview_SR}.

\begin{table}
\tbl{Summary table of the sum rules existing in the literature. The table is taken from \cite{Gehrlein:2015ena}.}
{\begin{tabular}{c c c c c c c}
\toprule
Sum rule & References & $c_1$&$c_2$&$d$&$\Delta \chi_{13}$&$\Delta \chi_{23}$ \\
\colrule
	1& \cite{Barry:2010zk,Bazzocchi:2009da,Ding:2010pc,Ma:2005sha,Ma:2006wm,Honda:2008rs,Brahmachari:2008fn,Kang:2015xfa,SR1} &$1$&$1$&$1$&$\pi$&$\pi$\\
	2& \cite{SR2} &$1$&$2$&$1$&$\pi$&$\pi$\\
	3& \cite{Barry:2010zk,Ma:2005sha,Ma:2006wm,Honda:2008rs,Brahmachari:2008fn,Altarelli:2005yx,Chen:2009um,Chen:2009gy,Kang:2015xfa,SR3} &$1$&$2$&$1$&$\pi$&$0$\\
	4& \cite{SR4} &$1/2$&$1/2$&$1$&$\pi$&$\pi$\\
	5& \cite{SR5} &$\tfrac{2}{\sqrt{3}+1}$&$\tfrac{\sqrt{3}-1}{\sqrt{3}+1}$&$1$&$0$&$\pi$\\
	6& \cite{Barry:2010zk,Bazzocchi:2009da,Ding:2010pc,Cooper:2012bd,SR6,Gehrlein:2014wda } &$1$&$1$&$-1$&$\pi$&$\pi$\\
	7& \cite{Barry:2010zk,Altarelli:2005yx,Chen:2009um,Chen:2009gy,Altarelli:2009kr,SR7,Altarelli:2008bg} &$1$&$2$&$-1$&$\pi$&$0$\\
	8& \cite{SR8} &$1$&$2$&$-1$&$0$&$\pi$\\
	9& \cite{SR9} &$1$&$2$&$-1$&$\pi$&$\pi/2,3\pi/2$\\
	10& \cite{SR10,Hirsch:2008rp} &$1$&$2$&$1/2$&$\pi,0,\pi/2$ &$0,\pi,\pi/2$ \\
	11& \cite{SR11} &$1/3$&$1$&$1/2$&$\pi$&$0$\\
	12& \cite{SR12} &$1/2$&$1/2$&$-1/2$&$\pi$&$\pi$\\
	\botrule
\end{tabular}}
\label{tab:overview_SR}
\end{table}

The sum rules had been known before, for recent overviews, see, e.g.~\cite{Barry:2010yk,SR11,King:2013psa}.
But so far it was not studied how the predictions of the mass sum rules are
affected by renormalisation group equation (RGE) corrections which we did in \cite{Gehrlein:2015ena}.
We do not want to discuss here the numerical results, for which the reader is
kindly referred to the original publication \cite{Gehrlein:2015ena}. Instead
we will discuss some analytical estimates wich show that one of the
most important qualitative features of the sum rules is robust under RGE
corrections.

The typical size of RGE corrections in the minimal supersymmetric extension
of the Standard Model (MSSM) can be estimated to be \cite{Antusch:2003kp}
\begin{align}
\delta \theta_{ij} &\sim 10^{-6} (1 + \tan^2 \beta) \frac{m^2}{\Delta m^2} \;,\\
\delta \phi_{i} &\sim 10^{-6} (1 + \tan^2 \beta) \frac{m^2}{\Delta m^2} \;,\\
\delta m_i &\sim (\mathcal{O}(1) + 10^{-6} (1 + \tan^2 \beta) ) m_i \;,
\end{align}
where $m^2$ and $\Delta m^2$ stands for the corresponding neutrino
masses and mass squared differences.
In the Standard Model without supersymmetry there is no factor of $\tan \beta$ such
that the RGE corrections there are expected to be rather small. On the other
hand they can become quite sizeable in the MSSM for large $\tan \beta$ 
and a large neutrino mass scale. So in this case one might wonder if
the corrections are large enough to allow a neutrino mass ordering which
would be forbidden on tree level.

As an example we study sum rule 2, cf.~Table~\ref{tab:overview_SR} which
reads
\begin{equation}
m_1 \text{e}^{-\text{i } \phi_1} + 2 m_2 \text{e}^{-\text{i } \phi_2} - m_3 = 0 \;.
\end{equation}
The sum rules can as well be interpreted geometrically in the complex plane,
cf.~\cite{King:2013psa, Gehrlein:2015ena}, as a closed triangle. Then we find for one of its angles on tree level
for sum rule 2 and inverted ordering
\begin{equation}
 \cos \alpha^{\text{tree}} = \frac{m_1^2 - 4 m_2^2 - m_3^2}{4 m_2 m_3} < - \frac{1}{4} \left( 3 \frac{m_2^2}{m_3^2} + 1 \right) < -1 \;. 
\end{equation}
Since the modulus of the cosine of an angle in a triangle is restricted to be
smaller than one we see that inverted ordering cannot be realised on tree
level. Two sides of the triangle are too short compared to the third side
to close the triangle.

An approximation of the RGE effects on $\cos \alpha$ gives in the MSSM
where the effect is expected to be sizeable
\begin{equation}
\delta(\cos \alpha)^{\text{RGE}} \approx - \underbrace{\frac{C y_\tau^2}{192 \pi^2}}_{>0}
\underbrace{\frac{2.8 m_1^2 - 0.4 m_2^2 + 0.1 m_3^2}{m_2 m_3}}_{>0} \underbrace{\log \frac{M_S}{M_Z}}_{>0} < 0 \;.
\end{equation}
For details on the derivation and the notation, please see the original paper \cite{Gehrlein:2015ena}.
The important thing is, that the correction is negative and hence points in the wrong
direction. This statement is true for most of the sum rules in the overwhelming part of
the parameter space. For the very few cases where the sign is correct one would
still need very extreme parameter choices, for instance $\tan \beta > 500$ or $m_1 > 1$~eV,
to reconstitute the forbidden ordering by RGE effects.

Note that this is particular to RGE corrections. Other kind of corrections, like  higher-dimensional
operators, flavon misalignment and so on, have in principle an arbitrary sign and might reconstitute
forbidden orderings in a plausible parameter range. But this is subject of another study in progress
\cite{Gehrlein:2016xyz}

\section{A Powerful Example}

As we have seen there are two kinds of sum rules. On the one hand there are sum rules
predicting the Dirac CP violating phase from the mixing angles and on the other hand
there are sum rules predicting the Majorana phases from the neutrino masses.

Now we want to briefly show how powerful it can be if a model incorporates both
sum rules making the model extremely predictive. As an example we want to take
again the $SU(5) \times A_5$ model \cite{Gehrlein:2014wda}. In this model
the neutrino sector exhibited golden ratio type A mixing \cite{Kajiyama:2007gx}
\begin{equation}
\tan \theta_{12}^\nu = \frac{2}{1+ \sqrt{5}} \;,  \quad \theta_{23}^\nu = \frac{\pi}{4} \; , \quad \theta_{13}^\nu = 0 \;.
\end{equation}
Due to its GUT nature and the use of alternative new GUT relations \cite{Antusch:2009gu}
there is only one sizeable mixing angle in the charged lepton sector for the
first two generations
\begin{equation}
\theta_{12}^e \approx \theta_C \Rightarrow \theta_{13} \approx \frac{\theta_C}{\sqrt{2}} \approx 9^\circ \;,
\end{equation}
where $\theta_C \approx 12^\circ$ is the Cabibbo angle. Related to this is
the mixing sum rule
\begin{equation}
 \theta_{12} \approx \theta_{12}^\nu + \theta_C \cos \delta \;.
\end{equation}
As mentioned above the model also has the mass sum rule
\begin{equation}
 \frac{\text{e}^{\text{i } \phi_1}}{m_1} + \frac{\text{e}^{\text{i } \phi_2}}{m_2} = \frac{1}{m_3}  \;.
\end{equation}
An additional important information is that to get the correct
GUT Yukawa relations a rather large $\tan \beta \gtrsim 30$
which implies that RGE corrections can be sizeable depending
on the neutrino mass scale.

From the mass sum rule we can estimate that
\begin{align}
 0.011 \text{ eV} & \lesssim m_1 && \text{for normal odering,} \\
 0.028 \text{ eV} & \lesssim m_3 \lesssim 0.454 \text{ eV} && \text{for inverted ordering.} 
\end{align}
Especially the lower bound for the neutrino masses is interesting
since it will give a lower bound on the running effects.

From the mixing sum rule we find an allowed range for $\theta_{12}$
at the high scale $M_S$
\begin{equation}
24^\circ \lesssim \theta_{12} (M_S) \lesssim 39^\circ \;.
\end{equation}
From the lower bound on $\tan \beta$ and the neutrino mass scale
we can now estimate the high scale range of $\theta_{12}$ by evolving
the low energy 3$\sigma$ range up to the high scale and we find
\begin{align}
 \theta_{12} (M_S) &\lesssim 33.5^\circ && \text{for normal odering,} \\
 \theta_{12} (M_S) &\lesssim 5.7^\circ && \text{for inverted ordering.} 
\end{align}
By comparing the ranges derived from the model at the high scale
and the RGE evolved ranges from the low scale we see that
the inverted ordering is by far excluded. This would not have been
the case, if there had been no mixing angle sum rule
on top of the mass sum rule. Hence, the combination of two kinds
of sum rules can be extremely powerful.

\section{Summary and Conclusions}

Sum rules are a common feature in flavour models and they can appear
in two different incarnations, either as mixing sum rules which relate the Dirac
CP phase to the mixing angles and/or as neutrino mass sum rules relating
the neutrino masses to the Majorana phases.

Both of these sum rules are easily testable in the near future. The Dirac CP phase
might be measured already very soon while mass sum rules are more difficult to test.
The most promising observable to distinguish between different sum rules is the
effective neutrino mass which can be determined from neutrinoless double beta decay
where no precision measurement is foreseeable in the near future. But even there by
simply measuring the ordering of the neutrino masses and the absolute neutrino mass
scale several of the sum rules and hence a lot of models would be immediately excluded.

Nevertheless, the testability of a high scale model should always be questioned taking
renormalisation group effects into account which can be very large in the neutrino sector
and alter the predictions at low energies. In fact, as we have demonstrated RGE corrections
have been crucial to understand why the inverted ordering of the neutrino masses was excluded
in our example $SU(5) \times A_5$ model.

Finally, to really understand flavour at the high energy scale we will have to test the TeV scale
extensively at colliders to discover the mechanism which makes the Higgs boson mass natural
or to abandon the notion of naturalness altogether.

\section*{Acknowledgments}

The author would like to thank the organisers of the conference
for the kind invitation and pleasant atmosphere. He is  supported  by 
BMBF  under  contract no.\ 05H12VKF and would like to thank
the Indonesian Institute of Science (LIPI) and KEkini for kind hospitality during which
this proceedings were finished.

\bibliographystyle{ws-rv-van}

\end{document}